\documentclass[12pt]{article}  
\usepackage{amsmath,amsfonts,amssymb}

\usepackage{hyperref}

\font\tenfrak=eufm10
\font\sevenfrak=eufm7
\font\fivefrak=eufm5
\newfam\frakfam  \textfont\frakfam=\tenfrak
  \font\tendl=msbm10  scaled \magstep1
\font\sevendl=msbm7 scaled \magstep1
\font\fivedl=msbm5 scaled \magstep1
\newfam\dlfam  
\textfont\dlfam=\tendl \scriptfont\dlfam=\sevendl
\scriptscriptfont\dlfam=\fivedl\scriptfont\frakfam=\sevenfrak  \scriptscriptfont
\frakfam=\fivefrak
\input{cyracc.def}

\newcommand{\nodo}[1]{}
\def\mat22#1#2#3#4{ \left(\begin{array}{cc} #1 & #2\\ #3 & #4
\end{array}\right) }

\newcounter{point}

\def\alg2{{\rm alg}_2}

\begin{document}

\title{Why blockchain and smart contracts need semantic descriptions}

\author{Zoran \v{S}koda}

\date{} 

\maketitle

\begin{abstract}
  We argue that there is a hierarchy of levels describing to
  that particular level relevant features of reality behind
  the content and behavior of blockchain and smart contracts
  in their realistic deployment.
  Choice, design, audit and legal control of these systems could
  be more informed, easier and raised to a higher level,
  if research on foundations of these descriptions develops
  and sets the formalisms, tools and standards for such descriptions.
\end{abstract}

\newif\ifzsnoco
\zsnocotrue 

\section{Introduction}
Blockchains and smart contracting platforms are usually presented either in
  rather algorithmic (or architectural) manner, emphasizing technical
  implementation or as highly idealized and simplified devices for highly
  controllable application scenarios. Surely, there is some common awareness
  of fallability or misuse of some of intended mechanisms which
  is typically attributed to a rather vague
  dependence on surrounding ``ecosystem''.
  While these pictures are suitable for some engineerings tasks, or
  as a first introduction to users, 
  they are typically insufficient for critical analysis
  needed by those initiating or systematically controlling their deployment,
  say policy makers, platform choosers, smart contracting platform architects.
  If this technology becomes prominent in real life clarity of recommendation
  will become a daily need. To give a metaphore, if we need to use a vehicle
  it is not sufficient to take some vehicle and to know how
  to drive one vehicle, but the vehicle and driving have to be appropriate,
  should it be a bike, boat, jeep, tank, motorcycle, an airplane or dirigible.
  The features of blockchains and smart contracts are even more diverse while
  much more opaque to observe than in the vehicle metaphore.

  It will take us a while to describe what we mean by {\em semantic description} and the account below is somewhat personal. In different areas, like philosophy, linguistics, computer science and logic, semantics refers to a slightly different concept or a subdiscipline. Our search is maybe closest to the usage in
  programming languages context. In early times of programming,
  code was written mainly ad hoc using whatever tricks and tools, which
  gradually included sophisticated compilers, debuggers, task scheduling
  in engineering teams and so on. Regardless sophistication, a question arose
  if two different implementations really operate the same,
  is the code verifiably correct and so on. Therefore the meaning
  and/or operational content should in particular be in some sense the same,
  whatever correct implementation is. Hence one had to answer the question what  is the operational content, what is the effect of the code. Different approaches like denotational and operational semantics, agent model etc. are devised to define properly and model this content. 

  Blockchains and more general distributed ledgers are distributed systems physically realized by a network of computers as nodes, whose making involves concurrency, consensus and possibly adversaries in the system. Hence its behaviour is not fully predictable. However, the genius of the invention is that incentives to parties operating in the system make certain aspects of handling the data which enter the ledger, which includes smart contracts, well behaved, reliable, replicable and so on to large extent. For some consensus mechanisms, very recent data on the ledger may be only tentative as choice of recognized fork takes some latency time and leads to abandoning some very recent blocks. The data on the ledger are processed and checked multiple times by varios nodes, leading to inefficiencies in addition to the inefficiencies which come from the computational price of consensus mechanism, say proof of work. For reasons of scalability and price, some data which logically have direct meaning to the algorithms in
  smart contracts on chain (= written in the ledger) are stored off chain, and only the handles (hashes or other identifiers) are stored on the chain. The hashes on the chain make verifying that the off chain data are correct cryptographic provable, however additional incentives are needed to mitagate the risk of
  unavailability of these data to complete the information needeed in the algorithms of smart contracts on chain. Thus, the availability of off chain data is one semantic characteristic of these data, which makes reality of smart programs very different then when the data are completely on chain. This is still engineering level, recognized in blockchain community.

  Much deeper problems happen when we include in the consideration meaning of the data processed by smart contracts in real world, like tokenized representations of assets, witness statements on achieved milestones in real world, timing information and so on. Clear picture is of ultimate importance for security of smart contracts: can contracts be explored by third parties, how to resolve conflicts, what is the risk analysis for assets and finally how should courts analyze and judge about the intended content and resolve the abuse, conflicts or technical glitches in procedures involving smart contracts. Some of the assets,
  like cryptocurrencies have their reality on chain and analysis of risks is limited to on chain plus possibly some off chain related second layer platform data. Many such analyses exist already. Next level is considering witness mechanisms. There are several cryptographic mechanisms how information about real
  world is verified on blockchain. First of all, smart contract can name holders of cryptographic keys of some blockchain addresses as verified witnesses who can enter information into a smart contract and this information is considered
  relevant, that is nontrivially influences the computation. These are usually
  called oracles. In a variant, oracle addresses are not predetermined but anybody who puts certain stake can be an oracle, and information is weighted over a pool of oracles who are rewarded for information but whose part of a stake is taken if the oracle's prediction is incorrect in the sense that it is far from the information given by majority of other oracles. This system is more in the spirit of decentralization than in advance certified oracles, but it requires handling the oracle market, which needs volume, good balance of incentives and may suffer from usual ideas of gaming the system which happen in rating and popularity systems. Oracle mechanisms may be combined with more involved cryptographic proofs from off chain world. That means that on chain, there is a preset verification test which can verify the cryptographic proof.

  Let us also mention the timing. Unlike usual programs which may run
  continuously and interact with surrounding signals during that execution,
  smart contracts run only when they are called, and the execution finishes
  within one block. There can be complementary off chain device which may
  be able to initiate a call to a program and take care to do it at right time.
  The blockchain itself has very rough timing information in the blocks,
  precise consensus on time is usually not present and it is difficult
  to incorporate beyond very rough precision. Of course, a smaft contract can
  itself call another smart contract and we can have a chain of executions.
  There are now variants, depending on blockchain in place. Either calls have
  to be executed also within the same block in which the original smart
  contract has been under execution or only a message has been written on
  chain by one smart contract to another and then this message will be
  interpreted as a call in the next block. The latter is useful if the
  platform is such that sharding has been extensively used, that is one
  divides the job between groups of miners, and then if a call has been made
  for a contract in a differeng shard it may be not feasible to execute
  within the same computing cycle but rather a request is made via
  a message which is on chain.

\section{Basics on what is semantics}

To setup our problem it is good to rethink of what a semantics can be. 
In different disciplines like philosphy, linguistics, computer science,
and logic, by semantics we mean somewhat different things. Maybe we
could say that a common ground is that the situation considered in most
settings is that we have some sort of reality and a presentation
of this reality in somewhat symbolic terms, say a thought, language construct,
formal language construct, computer program. There have been volumes
of philosophical discussion on the character of the correspondence between
the presentation and the reality. In the case of human language a convention
sets up what is the standard interpretation of some utterance. According to
Hirsch~\cite{Hirsch},
however the only consistent interpretation of human texts is the
one which tries to find which meaning has been intended by the creator
of the text; other interpretations are possible to produce but do not
truly pertain to the task of text interpretation. Thus, the linguistic
conventions are not exhaustive factors in human communication. Our
experience brings awareness that while our thoughts can be arbitrary,
interaction with other beings and our existence make that our arbitrary
will is useless in comparison to following patters in appearance of
new reality from known reality. This brings distinguishing some facts
as more real, truthful to others, and some sequences of thoughts as
regularly bring new things again as real, hence conclusions of truthful
statements. The reasonings in that endeavor boil down to logical
reasoning, not only classical, but if reality is understood in various
senses of useful experience, including conditional, hypothetical,
contractual and so on leading to various kinds of logic. These reasoning
chains are as a rule finite. Our experience can simply not confirm
others, except indirectly as artifacts of finite ones, though reality
also may have non finite ones which we do not comprehend as part of
our experience. Formalization of such logical systems,
and axiomatic bases of knowledge can again
be presented in terms of finite presentations by states in a computer
or sequences (or nonlinear utterances) of symbols. Formalization
then has two aspects, one is the system of utterances itself (basic symbols
and the syntax use to organize more complicated systems from basic
symbols), another is the intended representation. One can however
have another system for which, using the same rules of logical reasonings,
one can have valid reasoning for a different intended reality.
In mathematical logic these are called models.

In the study of computing, programs are producing reality rather than
proofs, at least at first inspection. This reality can be abstracted to
operations which happen with data (operational semantics), or we can go
toward assigning denotations to programming language constructs which
will correspond again to the reality of program execution. This is very
vague, but enough for rough comparison that we are talking about the same
subject. Again the program is just a symbolic presentation which has no
meaning outside of the system. Knowing how to present semantics of
a computer language is extremely important. For example, if we
change a platform and redo programming some critical system for this
platform in the langauge suitable for this platform do we get the same ?
Do the outputs of compiling the same code by two different compilers
give the same output ? Usually, computer languages do not have fully
determined semantics by their design. Namely there are situations
where an undefined behavior is possible (say choices). Some languages
are designed so that their semantics is very predictable, especially
functional languages what makes them more relevant in recent years.
These languages are designed having
in mind semantics of type theory~\cite{ChlipalaGirard},
a flavor of a logical system which came out of philosophy and logic
but has wide applications nowdays. Programs in this world can be viewed
as producing proofs of truth for some logical deductions. Useful
theoretical connections to category theory are widely explored. 

Regarding that the early languages for smart contracts (say, Solidity)
are rather ad hoc written without advanced programmin language theory properly
adapted to blockchain, some systems nowdays produce smart contracts
from functional languages, then there is an intermediate presentation
which formalizes some level of semantics, which is prone to better
analysis, for verification, security and other reasons. Then,
from the intermediate representation one produces automatically,
in a way similar to code generation in compilers, an actual program
in old fashioned smart contract language. Optimization reasons
suggest nowdays that the operational content should be done in
low level bytecode virtual machines and Webassembly VM seem to be a
tendency, complemented by system interface (set of standardized
system calls from VM). This way modularity of interaction between
the execution in VM and other aspects of blockchain and of off chain
reality is recognized, but not sufficiently modeled at abstract level.

\section{Assets on blockchain}
  
  Handling off chain assets on blockchain requires a representation of the assets on blockchain, most usually, but not exclusively, in tokenized form. This is in a complete analogy to an auction. In an auction, parties sign up before an auction and they are preverified for ability to pay. During the auction no real money is involved, certified party enters purchase by a promise statement. After the auction, off auction mechanisms are involved to collect the money promised. Similarly, an off chain asset is represented on the blockchain showing the representation of the asset and a blockchain address of the owner (the owner may be even some other smart contract). Now a number of issues come into
  play. First of all, with cryptocurrencies a double spending problem is solved on chain but the existence and possible multiple representations in different systems of a real asset can be regulated only off chain, in real world and with help of a legal system. Say, if some jurisdiction decides that all real estate in this jurisdiction is represented on some blockchain as a primary place of verification, then the issuer of real world certificate to this property, and that is the jurisdiction itself is the default place of a resolution of a conflict. Namely if some property is sold elsewhere and not on the authorized blockchain it should be considered legally not binding, and in most cases cheating or criminal misappropriation.

  Most proponents of smart contracts in past put the emphasis on variants of smart contracts which contain also the legal statements written on blockchain, describing the smart contract in legal terms. These so called
  Ricardian contracts are hence a pair of a real contract and a smart contract.
  This is typically considered as safe if the legal contract described the smart contract sufficiently precisely. One can always have a less verbose real contract which just says, accept whatever this code gives, making the promise to respect the outcome of the contract, but this opens the possibility that if one side of a contract is technically superior there can be hidden advantages in
  a contract designed and offered by such a side. This is very similar to the notion of business intelligence, but it is far ore opaque. In the case of business intelligence, say, a bank user opens an account or more complicated financial device and signs some document with the bank. The account is handled by a program set up by the bank, and has many defaults hwihc are typically not specified in the account and it is often not fully in agreement with all exceptions which exist elsewhere in the law or are tacit assumptions of the user when signing the contract about the financial device. If a user sees that something worked out different from the intution, the bank representative will usually say this is the way our computer handles this situation, I am sorry for this, but this is the business rule. It appears as the business logic within the bank is above the external contract between the bank and the user and brings additional rules which are not in the contract. This is an abuse of power by the bank per excellence. Now if the business logic is replaced by procedures on the blockchain
  then in principle, the user can check the correctness and so on and does not
  rely on hidden third party software and unchecked computation within the bank (we are neglecting the problem of account privacy here, as we consider a different aspect). But still the user can not have a way to audit the system without
  third party services. And even then, if we come to the court, is there a
  decription of the reality behind the smart contract which makes the implications of the contract clear enough for a court resolution. The fact that the code is visible, at least, and the execution deterministic (up to factors like the content of oracle messages, execution timing etc.), makes it less attractive to
  cheating. The collection of processing data is simpler and achieving to a
  conclusion is simpler as long as parties do not dispute the content of the
  smart contract but only following or not in real world the consequences
  of the actions of smart contracts on symbolic representations of real world assets. 

  In traditional systems there are some mechanisms of handling the inequality
  in comprehension of contracts. This rules are hard to enforce in practice
  as it is hard to prove both the intention to cheat, and miscomprehension
  of a side in a contract. In situations like selling a material item
  over the internet, there is a tendency in law that a customer (the weak
  party) should read more details of the contract if it is expected that
  (s)he is less familiar. For example, if you already used the service or
  if you are profiled as highly educated the system may simplify the procedure
  in which you are forced to check for more details. This is one of the prime
  cases belonging to the subject of the personalized law~\cite{Busch,BuschdF}. 
  
\section{Hierarchies}   

Which parts of the blockchain reality or smart contract reality
we should model? It is important for a problem (say a security analysis,
risk analysis, legal analysis, porting to another platform, refactoring
code) to be able to limit the content to a closed system using only
idalizations abstracting from phenomena which are either irrelevant or
could be separately modelled.

The very basic one is off chain and on chain in the strict sense. Hence we
do not imply the data or programs which are represented by hashes on the
blockchain or which are initiated by change of flags on the blockchain.
Even this basic reality of transactions in blocks may have variants. One
is to include or not the temporary forks which may be dispenses later.
Another is how deeply we go into the semantic of data within the on chain
transactions. For example, do we take into account the data as programs
(smart contracts) or not.

Now say, we consider smart contracts as programs. These programs do
not execute when we want. Namely, in each block miners include some
transactions, these transactions may be calls to smart contracts
from the account or from other smart contracts together with a trace
these smart contracts make on chain, that is the part of the results of
execution which concern the global variables, those whose value is written
on chain. There are incentives that the miners include certain transactions
but it may be that a transaction is delayed several blocks just because it
happens that the miners selected other transactions as more interesting
for inclusion into blocks. 

Typically most variables of
a smart contract are off chain. Do we work in idealization in which these
data are available or we take into account conditional semantics
in which these data are not secured ? 

There are two very different problems we want to describe and this is
in a complete analogy with programming languages. In the latter case, one
is the semantics of a programming language in the sense of a model
how to model all programs in that programming language. In a research
people study aspect by aspect of a programming language when doing this.
Another is the semantic representation of an instance, a particular program.  
This assumes producing tools, used say for verification or so. The two
realities are not much different in that case. In the case of smart contracts,
however, studying the expressive content of say a smart contracting language
is usually at a higher level of abstraction than when we study the content of a particular system of smart contract. In the latter case, the semantics of
data used, may be considered in a more speciic way, if we know the context.
So, without a context, say a variable of type asset is sort of class which
has some features, for example it is desirable that the language does not
allow making copies but only transfer of the asset values. Such things
were already used in the design of a more advanced languages for smart contracts, say Libra's Move language. In a context, we may have an information which
is not part of the semantics of the smart contracting language. For example,
we may know that some oracle's addresses are trusted and others are semireliable or behave as selfish agents collecting rewards in a way regulared by given smart contracts. Thus the semantics of the programming language for smaft contract, which is already specific is less expressive than the semantics needed to
model some smart contract in the context. From the point of view of research,
although the real life contexts are limitless, we indeed have some typical
colorings of the data which mainly center around the notions of availability,
trust and timeliness.

The whole situation is also prone to cryptographic factors. Users can loose
cryptographic keys, keys can be compromised and the information in various
channels of execution can or can not see information in other channels
due encryption, off chain execution mechanisms and so on. This is the usual
situation in distributive systems. Concurrency and cryptography has been
studied extensively using agent systems, process algebra, modal logic and
its modal theory, some aspects of which like Kripke models, bisimulation etc.
appear also in categorical treatments. 

Another aspect is when moving part of the computation to a state channel of
chain. Namely, for scalability purposes allowing wider applications, or
for privacy or other reasons, one can sometimes move some computation off chain
and expect some guarantees that the computation has been correct. It can be a
cryptographic proof, or else, all interested parties for a subset of smart
contracts instances, may be present in side channel with the same computational
rules as on the main chain and if a computation in some step of the channel does
not follow the rules a party can challenge the execution on the main chain. The
steps in between are digitally signed so the complaining party has the proof that
the block before the challened block happened. Then a verfication game can
resolve the dispute at the level of the main chain. Now, the execution in
state channel uses some data from the main chain which were withdrawn to the
channel, for example, some assets can be frosen on the main chain and used for
the computation in the channel and then at the moment of closing he channel,
many blocks later, the redistributed assets are returned to the main chain.
This phenomenon of temporary passing of a part of the state to a channel is
usually regulated by the rules of another system of smart contracts which
establishes the mechanism of setting up the channel. Thus this is
not a characteristic of a blockchain itself, but of a smart contracting
platform which includes the system of contracts regulating the issue. Of course,
such mechanisms can be included potentially in the basic protocol
of some blockchain and the instances will just require signing up
to an instantiation of such a channel. However this is not obvious as it
is not clear which kind of variables other than most simple asset types
can figure out in such channeling procedures. 

It is an open question how to make these potential models compatible with the
reality of the contract legislation, which should adapt to smart contracts as
a variant with classica issues like incompleteness of contracts,
conflict resolution etc.~\cite{MacMillan}

\end{document}